# Hardware Implementation of a Polar Code-based Public Key Cryptosystem

Reza Hooshmand [*], Farhad Naserizadeh, and Jalil Mazloum

Department of Electrical Engineering

Shahid Sattari Aeronautical University of Science and Technology, Tehran, Iran

*Corresponding Author Email: rhooshmand@ssau.ac.ir

**Abstract**: In recent years, there have been many studies on quantum computing and the construction of quantum computers which are capable of breaking conventional number theory-based public key cryptosystems. Therefore, in the not-too-distant future, we need the public key cryptosystems that withstand against the attacks executed by quantum computers, so-called post-quantum cryptosystems. A public key cryptosystem based on polar codes (PKC-PC) has recently been introduced whose security depends on the difficulty of solving the general decoding problem of polar code. In this paper, we first implement the encryption, key generation and decryption algorithms of PKC-PC on Raspberry Pi3. Then, to evaluate its performance, we have measured several related parameters such as execution time, energy consumption, memory consumption and CPU utilization. All these metrics are investigated for encryption/decryption algorithms of PKC-PC with various parameters of polar codes. In the next step, the investigated parameters are compared to the implemented McEliece public key cryptosystem. Analyses of such results show that the execution time of encryption/decryption as well as the energy and memory consumption of PKC-PC is shorter than the McEliece cryptosystem.

## 1. Introduction

Nowadays, the security of some common public key cryptosystems, such as RSA [1] and Elgamal [2], depends on several number theory problems, e.g. integer decomposition and discrete logarithm problems. In 1994, Peter Shor shows the possibility of solving such number theory based problems using quantum computers in polynomial time [3]. Therefore, public key cryptosystems based on these types of problems will be vulnerable by constructing the quantum computers with adequate computing power. This means that the other suitable cryptosystems that are secure against such quantum computer-based attacks should replace the conventional public key cryptosystems. Such cryptosystems that can oppose against quantum computer-based attacks are called post-quantum cryptosystems. Code-based cryptosystems are known as a class of post-quantum cryptosystems whose security is based on the difficulty of problems related to channel coding, e.g. general decoding problem and syndrome decoding problem [4]. The first code-based public key cryptosystem was introduced by McEliece [5], whose security is based on the difficulty of solving general decoding problem for Goppa codes [6]. Unlike conventional public key cryptosystems, the McEliece cryptosystem is resistant against attacks based on quantum computers and also has a good level of security against conventional attacks [7]. However, due to increasing the computing power of computers as well as the introduction of optimized attacks, the main parameters of this cryptosystem need to be updated. One of the most important advantages of McEliece cryptosystem is that it has very fast encryption/decryption algorithms. However, it has several disadvantages such as low data rates and large key length compared to the conventional public key cryptosystems [8]. So far, several solutions and methods have been presented in the field of analysis and improvement of McEliece cryptosystem. For instance, exploiting an appropriate channel codes such as polar codes [9] in its structure and introducing McEliece-like cryptosystems is a proper way to improve this cryptosystem. Improving the efficiency level of cryptosystem such as increasing the code rate, reducing the key length and increasing the speed are other effective upgrade techniques.

In recent years, polar codes are applied in various fields of cryptography such as physical layer encryption scheme [10], secure channel coding scheme [11] and secret key cryptosystem [12], identification scheme [13] and key encapsulation mechanism [14]. Hence, due to unique capabilities of polar codes, it is reasonable to apply such codes in the structure of public key cryptosystems. Recently, a public key cryptosystem based on polar code (PKC-PC) has been proposed [15, 16]. The security of such cryptosystem is based on the difficulty of solving the general decoding problem (GDP) [17]. In [16], it is shown that by exploiting the characteristics of polar codes and applying a proper method in the PKC-PC structure, the public key size, and private key size, and the computational complexity are reduced in compared to the McEliece cryptosystem. Also, PKC-PC has an acceptable security level against conventional attacks such as exhaustive search attack, key recovery attack and information set decoding attack. Furthermore, it is proved that the security of PKC-PC is decreased to solve NP-complete hard problems, i.e. polar parameterized syndrome decoding (PPSD) and polar parameterized codeword existence (PPCE). In the field of PKC-PC efficiency analysis, the decryption failure rate (DFR), the key size and the computational complexity are evaluated.

In this paper, the PKC-PC encryption/decryption is implemented on an embedded system, i.e. Raspberry Pi3 [18]. Then, in order to evaluate its efficiency level, several parameters such as runtime, energy consumption, memory consumption and CPU utilization have been examined. In the next step, these parameters are compared with the parameters of the McEliece public key cryptosystem implemented on the Raspberry Pi3 [19]. Then, to evaluate its performance, we have measured several related parameters such as execution time, energy consumption, memory consumption and CPU utilization. All these metrics are investigated for encryption/decryption algorithms of PKC-PC with various parameters of polar codes. In the next step, the investigated parameters are compared to the implemented McEliece public key cryptosystem in order to evaluate its efficiency level. Analyses of such results show that the execution time of encryption/decryption as well as the energy and memory consumption of PKC-PC is less than the McEliece cryptosystem. The residue of this paper is organized as follows. In Section 2, we give a brief description of polar codes. In Section 3, we completely illustrate the structure of the PKC-PC. The hardware implementation of PKC-PC is explained in Section 4. Moreover, we discuss the advantages of the implemented PKC-PC compared to the implemented McEliece cryptosystem. At last, Section 5 presents the conclusion of this work.

## 2. Polar Codes

Polar codes are a series of linear block codes with the capability of achieving the capacity of any symmetric Binary-input Discrete Memoryless Channel (B-DMC), e.g. Binary Erasure Channel (BEC). Let $W : \mathcal{X} \to \mathcal{Y}$ be a B-DMC with transfer probabilities of $\{W(y|x), x \in \mathcal{X}, y \in \mathcal{Y}\}$. First consider two parameters $I(W)$ and $Z(W)$ for a B-DMC $W$ as defined follows [20]

$$I(W) \triangleq \sum_{y \in \mathcal{Y}} \sum_{x \in \mathcal{X}} \frac{1}{2} W(y|x) \log \frac{W(y|x)}{\frac{1}{2}W(y|0)+\frac{1}{2}W(y|1)}, \quad (1)$$

$$Z(W) \triangleq \sum_{y \in \mathcal{Y}} \sqrt{W(y|0)W(y|1)}, \quad (2)$$

In fact, $I(W) \in [0,1]$ is named as the *capacity* of $W$ and therefore is used for evaluating the rate. Moreover, $Z(W) \in [0,1]$ is named as the *Bhattacharyya parameter* of $W$ and considered for measuring the channel reliability. If $W$ is a BEC with erasure probability $\epsilon$, denoted by BEC($\epsilon$), then $Z(W) = \epsilon$ and $I(W) = 1 - Z(W) = 1 - \epsilon$. Let $\{W_n^{(i)}: 1 \leq i \leq n\}$ be a set of *polarized* B-DMCs with indices '$i$' that can be performed by a phenomenon, named *channel polarization*, on the $n$ independent copies of $W$. The Bhattacharya parameter for each $W_n^{(i)}$ is denoted by $(W_n^{(i)})$ [20].

## 2.1. Constructing the Generator Matrix

Let us consider the polar code length $n = 2^m$, $m \geq 1$, the polar code dimension $k = nR$, the polar code rate $R < I(W)$ and $F = \begin{bmatrix} 1 & 0 \\ 1 & 1 \end{bmatrix}$. A $k \times n$ generator matrix of any $(n, k)$ polar code is generated by the help of following steps [20]:

1) Obtain the $n$-th kronecker product $G_n = F^{\otimes m}$ which gives an $n \times n$ matrix. Also, label its rows from top to bottom as $i = 1, 2, \cdots, n$.

2) For $k = 1, 2, 2^2, \cdots, 2^{n-1}$, compute the Bhattacharyya parameters of all $n$ bit-channels as $Z_n = (Z(W_n^{(i)}), 1 \leq i \leq n)$ by the help of the recursive relation (3) with initial value $Z(W_1^{(1)})$. If $W$ is BEC($\epsilon$), $Z(W_1^{(1)})$ is considered as $\epsilon$.

$$Z(W_{2k}^{(i)}) = \begin{cases} 2Z(W_k^{(i)}) - Z^2(W_k^{(i)}) & 1 \leq i \leq k \\ Z^2(W_k^{(i-k)}) & k + 1 \leq i \leq 2k \end{cases}. \quad (3)$$

3) Form a permutation $\pi_n = (i_1, i_2, \ldots, i_n)$ for $n$ bit-channel indices set $\mathcal{I}_n = \{1, 2, \cdots, n\}$ such that the inequality $Z(W_n^{(i_j)}) \leq Z(W_n^{(i_k)})$, $1 \leq j < k \leq n$ is satisfied.

4) Construct the $k$-element *information set* $\mathcal{A} \subset \mathcal{I}_n$ whose bit-channel indices correspond to $k$ leftmost indices of the permutation $\pi_n$, i.e., $i_1, \ldots, i_k$. Also, generate the $(n - k)$-element *frozen set* $\mathcal{A}^c \subset \mathcal{I}_n$ which is the complement of $\mathcal{A}$.

5) At last, obtain the generator matrix $G_\mathcal{A}$ by selecting $k$ rows of the matrix $G_n$ corresponding to the bit-channel indices of the information set $\mathcal{A}$. Also, generate the complement matrix $G_{\mathcal{A}^c}$ by selecting $(n - k)$ rows of the matrix $G_n$ corresponding to the bit-channel indices of the frozen set $\mathcal{A}^c$.

6) The transpose of the parity check matrix of polar codes $H_P^T$ is constructed by selecting the columns of $G_n$ with indices of the frozen set $\mathcal{A}^c$ [21].

One characteristic of matrix $G_n$ is that any its submatrix, including the array of elements $(G_{i,j})$, $i, j \in \mathcal{A}$, is a lower-triangular matrix which has ones on its diagonal, therefore it is nonsingular (invertible) [22].

## 2.2. Polar Encoding

For the polar codes of length $n$, an input vector $u = u_1^n = (u_\mathcal{A}, u_{\mathcal{A}^c})$ includes the *information vector* which is $k$-bit subvector $u_\mathcal{A} = (u_i, i \in \mathcal{A})$, and the *frozen vector* which is $(n-k)$-bit subvector $u_{\mathcal{A}^c} = (u_i, i \in \mathcal{A}^c)$. The input vector $u$ is polar encoded to $n$-bit codeword $x$ as follows:

$$x = u_\mathcal{A} G_\mathcal{A} + u_{\mathcal{A}^c} G_{\mathcal{A}^c}, \tag{4}$$

Therefore, the polar code rate is computed as follows [20],

$$R = |u_\mathcal{A}|/|x| = |\mathcal{A}|/n = k/n. \tag{5}$$

## 2.3. Successive Cancellation Decoding

The successive cancellation (SC) decoder architectures include the following four types: (i) butterfly-based architecture; (ii) pipelined tree architecture; (iii) line architecture and (iv) min-sum approximation [23]. In this paper, we have used the min-sum approximation for SC decoding. Let $y = y_1^n$ be the output vector of corresponding channel. The SC decoder [24], estimates the input vector $\hat{u} = \hat{u}_1^n$ by the help of the information set $\mathcal{A}$, the frozen vector $u_{\mathcal{A}^c}$, the channel outputs vector $y = y_1^n$ and the estimated input vector $\hat{u}_1^{i-1}, i = 1, 2, \cdots, n$, as follows:

$$\hat{u}_i = \begin{cases} u_i, & \text{if } i \in \mathcal{A}^c \\ h_i(y_1^n, \hat{u}_1^{i-1}) & \text{if } i \in \mathcal{A} \end{cases}, \tag{6}$$

where hard decision function $h_i: \mathcal{Y}^n \times \mathcal{X}^{i-1} \to \mathcal{X}, i \in \mathcal{A}$, is obtained as follows for all $y_1^n \in \mathcal{Y}^n, \hat{u}_1^{i-1} \in \mathcal{X}^{i-1}$:

$$h_i(y_1^n, \hat{u}_1^{i-1}) \triangleq \begin{cases} 0, & \text{if } \frac{W_n^{(i)}(y_1^n, \hat{u}_1^{i-1}|u_i=0)}{W_n^{(i)}(y_1^n, \hat{u}_1^{i-1}|u_i=1)} \geq 1 \\ 1, & \text{otherwise} \end{cases}. \tag{7}$$

It has been shown that for any given B-DMC $W$, the decoding failure rate for the channel outputs $y$ under SC decoding is upper bounded as follows:

$$P_e = P\{u \neq \hat{u}\} \leq \sum_{i \in \mathcal{A}} Z\left(W_n^{(i)}\right). \tag{8}$$

Also, reliable communication under SC decoding for any B-DMC $W$ is created when the following inequality is obtained [25]:

$$R < I(W) - n^{-1/\mu}, \tag{9}$$

where $\mu$ is named as *scaling exponent* and its value for BECs is $\mu_{\text{BEC}} \approx 3.627$. In this work, the maximum possible code rate satisfying the relation (9) is denoted by $R_0$.

## 3. The Construction of PKC-PC

Figure 1 shows the overall construction of PKC-PC [16]. As shown in this figure, the PKC-PC consists of key generation algorithm (KeyGen), encryption algorithm (Enc) and decryption algorithm (Dec). In the sequel of this section, we explain such algorithms in detail.

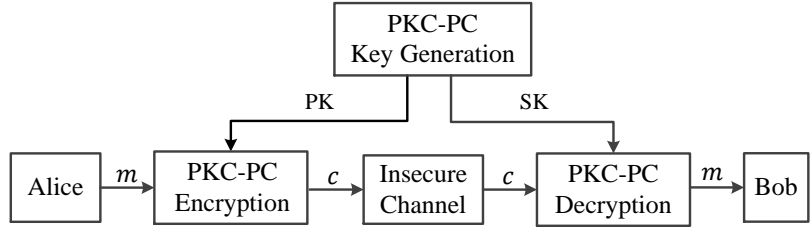

**Fig. 1.** The overall construction of PKC-PC.

### 3.1. PKC-PC Key Generation

The KeyGen of PKC-PC [16] generates public key (PK) and secret key (SK). The SK of PKC-PC is defined as the $\{\mathcal{A}^c(s), P\}$. Also, the PK of PKC-PC is considered as a $k \times (n-k)$ matrix $Q$ generated from $k \times n$ encryption matrix $G' = S^{-1}G_{\mathcal{A}(s)}P = [I_k|Q]$, where $S^{-1}$ and $I_k$ show the inverse of nonsingular matrix $S$ and the $k \times k$ identity submatrix, respectively. Algorithm 1 describes the KeyGen of PKC-PC and Figure 2 shows the KeyGen's flowchart [16].

*Algorithm 1: PKC-PC Key Generation (KeyGen)*

Data: the target security level $\lambda$

Input: the seed (s) of used CTR-DRBG.

Output: the public key (PK) $Q$ and the secret key (SK) $\mathcal{A}^c(s)$.

1. Given $\lambda$, fix the polar code parameters $n, k, R, R_0$.
2. Compute the $m$-th kronecker product $G_n = F^{\otimes m}$ as an $n \times n$ matrix and mark its rows as $i = 1, 2, \ldots, n$ from top to bottom.
3. For $k = 1, 2, 2^2, \cdots, 2^{m-1}$, with initial condition $Z\left(W_1^{(1)}\right)$, compute all $n$ bit-channels Bhattacharyya parameters as $Z\left(W_{2k}^{(i)}\right) = 2Z\left(W_k^{(i)}\right) - Z^2\left(W_k^{(i)}\right), 1 \leq i \leq k$ or $Z\left(W_{2k}^{(i)}\right) = Z^2\left(W_k^{(i-k)}\right), k+1 \leq i \leq 2k$.
4. Form a permutation $\pi_n = (i_1, i_2, \ldots, i_n)$ for $n$ bit-channel indices set $\mathcal{I}_n = \{1, 2, \cdots, n\}$ such that the inequality $Z\left(W_n^{(i_j)}\right) \leq Z\left(W_n^{(i_k)}\right), 1 \leq j < k \leq n$ is satisfied.

5. Generate a secret information set through a CTR-DRBG by which its $k$ indices are selected randomly from $nR_0$ leftmost indices in $\pi_n$, i.e., $i_1, \ldots, i_{nR_0}$.

6. Generate a secret generator matrix $G_{\mathcal{A}(s)}$ for an $(n, k)$ polar code by choosing $k$ rows of $G_n$ corresponding to the indices of $\mathcal{A}(s)$.

7. Generate $P_1$ as an $n \times k$ submatrix with $k$ '1's, each located in $j$-th, $j \in \mathcal{A}(s)$ row of its $k$ columns, respectively.

8. Generate $P_2$ as an $n \times (n-k)$ submatrix with $(n-k)$ '1's located in $j$-th, $j \in \mathcal{A}^c(s)$ row of its $(n-k)$ columns.

9. Set the concatenation of submatrices $P_1$ and $P_2$ as a permutation matrix $P = [P_1|P_2]$.

10. Generate the $k \times k$ nonsingular submatrix $S = (G_n)_{\mathcal{A}(s)\mathcal{A}(s)}$.

11. Generate the encryption matrix $G' = S^{-1}G_{\mathcal{A}(s)}P = S^{-1}G'' = [I_k|Q]$ and extract the submatrix $Q_{k \times (n-k)}$ as public key (PK).

12. Complement $\mathcal{A}(s)$ to obtain $\mathcal{A}^c(s)$ and set it as SK.

13. Return $\mathcal{A}^c(s)$ and $Q$.

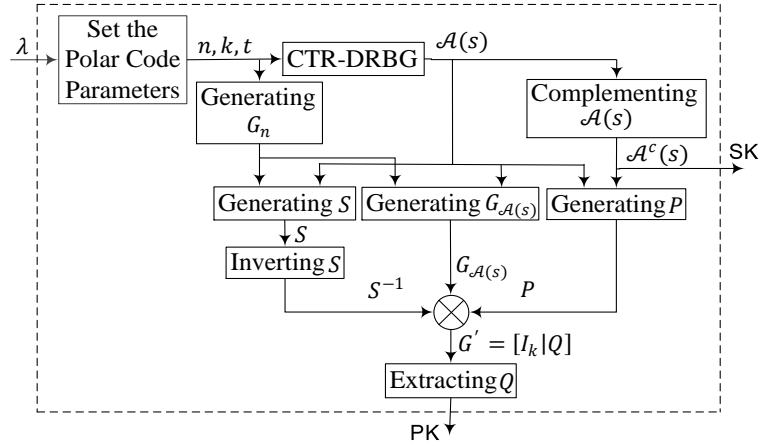

**Fig. 2:** The flowchart of the PKC-PC Key Generation.

### 3.2. PKC-PC Encryption

The encryption of PKC-PC [16] is executed as follows. Let us consider the message $m$ as a $1 \times k$ binary vector. To encrypt $m$, the sender selects $Q$ from the public directory and generates the encryption matrix $G' = [I_k|Q]$. Then, the sender generates the $1 \times n$ binary ciphertext vector as $c = mG' + e$, where $e$ is defined as a random binary error vector. In this case, the Hamming weight of error (perturbation) vector is less than the error correction capability of used polar code, i.e., $w_H(e) < t$. Algorithm 2 presents the Enc of PKC-PC and Figure 3 shows its corresponding flowchart [16].

*Algorithm 2: PKC-PC Encryption (Enc)*

Data: The parameters of used polar code ($n$, $k$ and $t$)

Input: The message $m$ and public key PK= $Q$

Output: The ciphertext $c$

1. $G' \coloneqq [I_k|Q]$
2. Construct the random perturbation vector $e$ with the Hamming weight $w_H(e) \leq t$
3. $c \coloneqq mG' + e$

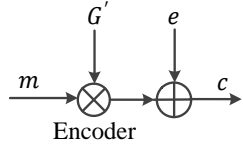

**Fig. 3:** The flowchart of the PKC-PC Encryption.

### 3.3. PKC-PC Decryption

The decryption of PKC-PC [16] is executed as follows. Let us consider the ciphertext $c$ as a $1 \times n$ binary vector. First, $c' = cP^{-1} = mS^{-1}G_{\mathcal{A}(s)} + eP^{-1}$ is obtained, where $P^{-1}$ is the inverse of the permutation matrix $P$. Since $P$ has the permutation property, we have $w_H(eP^{-1}) = w_H(e)$. Hence, $c' = c'^n_1$ is a codeword of used polar code previously selected and the perturbation vectors can be deleted with the help of SC decoding to obtain $mS^{-1}$. Since $u_{\mathcal{A}^c(s)}$ is a full-zero vector, the parameters $\{\mathcal{A}(s), c'\}$ are considered as the input of SC decoder (See Fig. 4). The input vector of polar encoder, $u = (u_{\mathcal{A}(s)}, u_{\mathcal{A}^c(s)}) = (mS^{-1}, 0)$, is evaluated with the help of the SC decoding as follows: (i) $\forall i \in \mathcal{A}^c(s)$, $\hat{u}_i = 0$; (ii) $\forall i \in \mathcal{A}(s)$, $\hat{u}_i = h_i(c'^n_1, \hat{u}^{i-1}_1)$. The hard decision function $h_i$ is defined as $\forall \frac{W^{(i)}_n(c'^n_1, \hat{u}^{i-1}_1 | u_i = 0)}{W^{(i)}_n(c'^n_1, \hat{u}^{i-1}_1 | u_i = 1)} \geq 1$, $h_i(c'^n_1, \hat{u}^{i-1}_1) = 0$; otherwise, $h_i(c'^n_1, \hat{u}^{i-1}_1) = 1$. In other words, if the index $i$ of $W^{(i)}_n$ is not in the $\mathcal{A}(s)$, then the decoder knows that $\hat{u}_i = u_i = 0$. When the SC decoder permutes $c'$ into $\hat{u} = \hat{u}^n_1$, the plaintext is recovered as $m = u_{\mathcal{A}(s)}S$. In the manner of existence of decryption failure, the receiver should request the sender to resend the ciphertext $c$ again. The PKC-PC decryption algorithm is presented in Algorithm 2 [16].

*Algorithm 2: PKC-PC Decryption (Dec)*

Data: The parameters of used polar code ($n$, $k$ and $t$)

Input: The ciphertext $c$ and $\mathsf{SK}=\{\mathcal{A}^c(s), P\}$

Output: The message $m$

1. $c' \coloneqq cP^{-1} \coloneqq mS^{-1}G_{\mathcal{A}(s)} + eP^{-1}$
2. $u \coloneqq \text{SC Decoding}\{c', \mathcal{A}(s)\}$
3. $u_{\mathcal{A}(s)} \coloneqq Msb_k(u)$
4. $m \coloneqq u_{\mathcal{A}(s)}S$

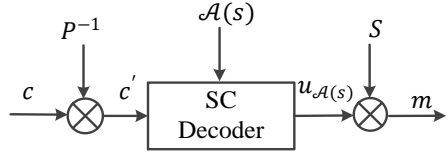

**Fig. 4:** The flowchart of the PKC-PC decryption.

Table 1 provides a comparison between the theoretical encryption/decryption complexity of PKC-PC and McEliece cryptosystem. Also, the required binary operation for one-bit encryption/decryption is computed. As shown in this table, assuming that $log n < t$, the complexity and required binary operation for encryption/decryption of PKC-PC is less than the McEliece cryptosystem.

**Table 1:** The theoretical comparison between PKC-PC and McEliece cryptosystem.

| Complexity<br>Cryptosystem | Encryption complexity | Required operation for one-bit encryption | Decryption complexity | Required operation for one-bit decryption |
|---|---|---|---|---|
| McEliece [5] | $\mathcal{O}(nk)$ | $n$ | $\mathcal{O}(n^2 + nt + k^2)$ | $(n^2 + nt + k^2)/k$ |
| PKC-PC [16] | $\mathcal{O}(k(n-k))$ | $n-k$ | $\mathcal{O}(n^2 + nlogn + k^2)$ | $(n^2 + nlogn + k^2)/k$ |

### 3.4. Security and Efficiency Analysis of PKC-PC

In [16], the efficiency level of PKC-PC is evaluated in terms of the decryption failure rate (DFR), the key length and the computational complexity. Moreover, formal security assessment of PKC-PC is presented in [16] by which the PKC-PC's security is reduced to solve the NP-complete problems, called as polar parameterized syndrome decoding (PPSD) and polar parameterized codeword existence (PPCE). Also, to investigate the practical security of PKC-PC, several attacks such as exhaustive key search attack, key-recovery attack and information set decoding (ISD) attack are considered. The investigation's results show that the security and efficiency of PKC-PC depend on various operators such as code length, code dimension and the Hamming weight of the error vector. In fact, these factors should be chosen in such a way that a suitable trade off will be performed between security and efficiency [16].

## 4. Hardware Implementation of PKC-PC

In this section, the hardware implementation of PKC-PC [16] is examined on the Raspberry Pi3. Also, the implementation results are compared with the implemented McEliece cryptosystem on Raspberry pi3 in [19]. In fact, the key generation, encryption and decryption algorithms of PKC-PC have been implemented separately with python programming language on Raspberry Pi3. The program written for each algorithm of PKC-PC is exactly in accordance with the inputs and outputs of such algorithm and the mathematical operations expressed in [16] are also implemented. The Raspberry Pi3 minicomputer has a 64-bit processor with a frequency of 2.1 GHz [18]. The main operating system for its launch is Raspbian which is based on Debian Linux. Figure 5 shows the Raspberry Pi3 hardware whose specifications are also described in Table 2 [19]. In the sequel of this section, the implementation results of PKC-PC are investigated and compared with such results in the implemented McEliece cryptosystem.

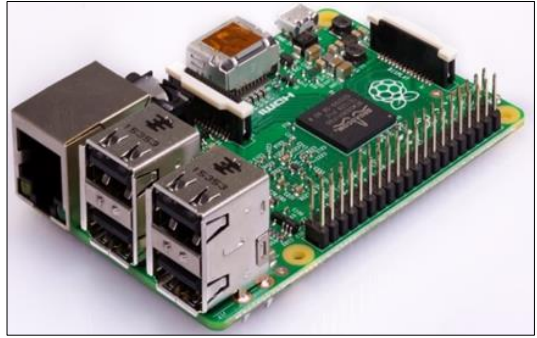

**Fig. 5:** Overview of Raspberry Pi3.

**Table 2:** Specifications of Raspberry Pi3.

| Device features | Specifications |
|---|---|
| SoC | Boardcom BCM2837 |
| CPU | 4*ARMCortex-A53,1.2GHz |
| RAM | 1GB LPDDR2 (900MHz) |
| Networking | 10/100 Ethernet, 2.4GHZ 802.11n wireless |
| Power supply | Micro USB Power Input |

## 4.1. Encryption/Decryption Execution Times

The encryption/decryption execution times of algorithms are one of the most important performance parameters for the PKC-PC. In this work, the essential point to have an accurate comparison between the encryption/decryption runtime of PKC-PC and McEliece cryptosystem is that the selected values of $n$ and $k$ for the implemented PKC-PC should be similar to the such parameters of implemented McEliece cryptosystem in [19]. The appropriate code rate is not the criterion of selecting $n$ and $k$ in such comparison.

Figures 6 and 7 show the comparison between the encryption/decryption execution time of PKC-PC and McEliece cryptosystem implemented on raspberry pi3 for $n = 256, 512, 1024$ and $k = 16, 8, 4$, i.e. with the public key size 512 Bytes. As shown in these figures, with respect to the given $n$, $k$ and $t$, the execution time of the encryption/decryption algorithms in the PKC-PC implemented on the raspberry pi3 is less than the same in the McEliece cryptosystem. In fact, the PKC-PC encryption algorithm is about 7 to 8 times faster than the McEliece encryption algorithm. However, due to the SC decoding calculations of PKC-PC decryption algorithm, the ratio of PKC-PC decryption speed to the McEliece decryption speed in not as much as the ratio of PKC-PC encryption speed to the McEliece encryption speed.

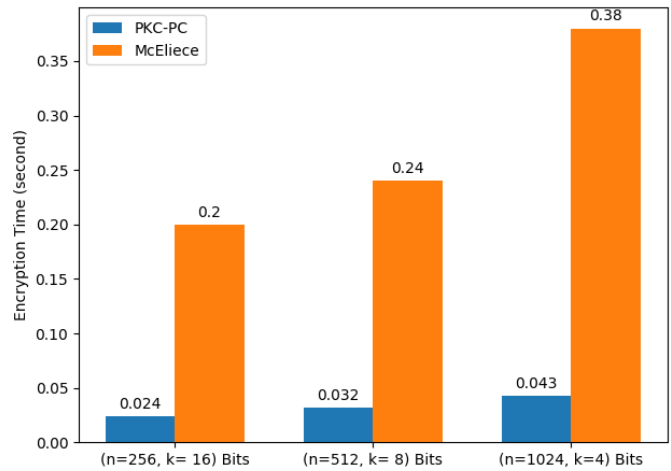

**Fig. 6:** Comparison between the encryption execution time (second) of PKC-PC and McEliece cryptosystem implemented on Raspberry Pi3 for $n = 256, 512, 1024$ and $k = 16, 8, 4$.

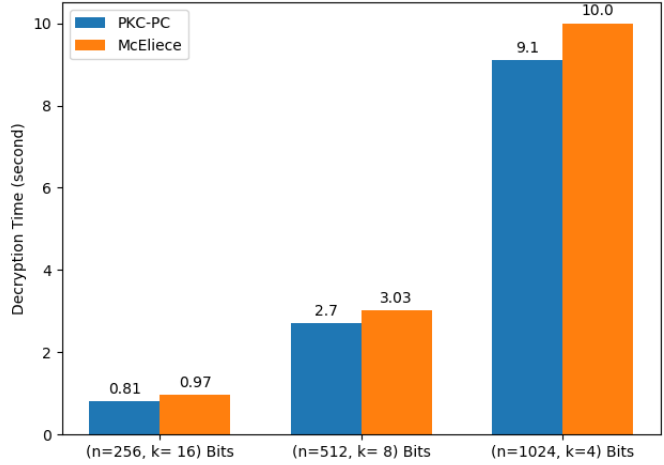

**Fig. 7:** Comparison between the decryption execution time (second) of PKC-PC and McEliece cryptosystem implemented on Raspberry Pi3 for $n = 256, 512, 1024$ and $k = 16, 8, 4$.

Figure 8 shows the execution time of PKC-PC encryption/decryption algorithms implemented on the Raspberry Pi3 for $n = 256, 512, 1024$ and $k = 192, 384, 768$, i.e. with code rate $R = 0.75$. In fact, the PKC-PC encryption/decryption algorithms in [16] is provided for the code rate $R = 0.75$. In this case, in addition to transmitting more information per encryption of each message block, a shorter public key length can be achieved compared to the lower code rate. For instance, the encryption/decryption times of PKC-PC for $(n, k) = (256, 192)$ polar code are computed as 0.076 and 1.24 seconds, respectively. Note that the implementation of successive cancellation (SC) decoding algorithm can be optimized [23] to reduce the execution time of the PKC-PC decryption algorithm.

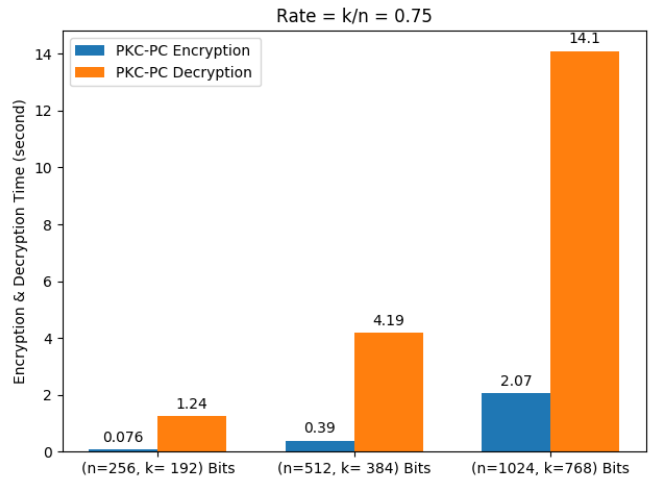

**Fig. 8:** The execution time (second) of PKC-PC encryption/decryption algorithms implemented on Raspberry Pi3 for $n = 256, 512, 1024$ and $k = 192, 384, 768$.

## 4.2. Memory Consumption

The memory consumption of executing the encryption and decryption algorithms for the PKC-PC is computed as $n(1 + k + t + 2n) + t^2$ and $n(3n + 2k + t + 1) + t^2 + k^2$, respectively [19]. In this case, $n$ is the polar code length, $k$ is the polar code dimension and $t$ is the error correction capability of used polar code. Figures 9 and 10 illustrate the obtained instances regarding the encryption/decryption memory consumption (kBytes) of implemented PKC-PC and McEliece cryptosystem for various message lengths and the same public key length, i.e. with the public key size 512 Bytes. For instance, for $n = 256$, $k = 16$, the memory consumptions of encryption algorithm for PKC-PC and McEliece cryptosystem are obtained as 14.43 and 18 kBytes, respectively. Also, for the same parameters, the memory consumptions of decryption algorithm for PKC-PC and McEliece cryptosystem are obtained as 21.42 and 23.73 kBytes, respectively. As shown in these figures, the PKC-PC encryption/decryption algorithms consume less memory compared to the McEliece cryptosystem for $n = 1024, 512, 256$ and $k = 4, 8, 16$, respectively.

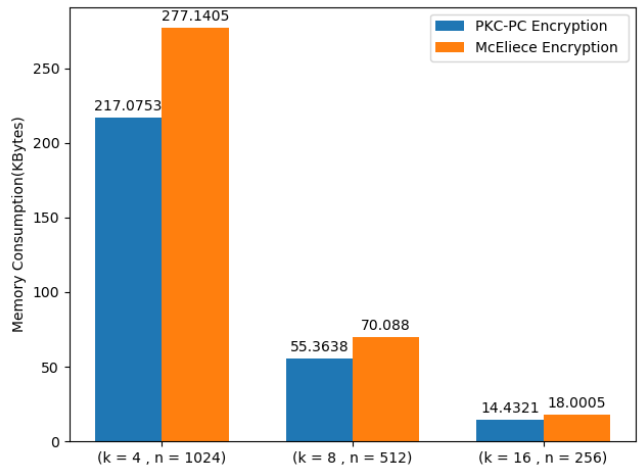

**Fig. 9:** The comparison between the encryption memory consumption (kBytes) of PKC-PC and McEliece cryptosystem implemented on Raspberry Pi3 for the same public key length 512 Bytes.

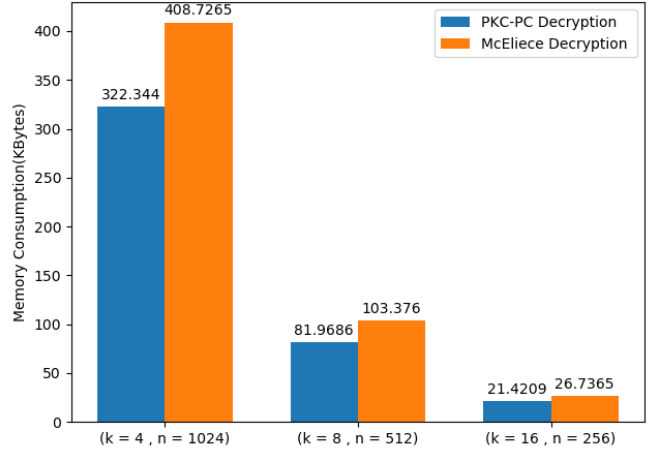

**Fig. 10:** The comparison between the decryption memory consumption (kBytes) of PKC-PC and McEliece cryptosystem implemented on Raspberry Pi3 for the same public key length 512 Bytes.

Figures 11 and 12 illustrate the comparison between the decryption memory consumption of PKC-PC and McEliece cryptosystem implemented on Raspberry Pi3 for the same plaintext length and various code lengths, i.e. various public key lengths. For instance, for $n = 64$, $k = 16$, the memory consumptions of encryption algorithm for PKC-PC and McEliece cryptosystem are obtained as 1.046 and 1.232 kBytes, respectively. Also, for the same parameters, the memory consumptions of decryption algorithm for PKC-PC and McEliece cryptosystem are obtained as 1.584 and 1.904 kBytes, respectively. As illustrated in these figures, for the same plaintext length, the encryption/decryption memory consumption of PKC-PC is less than the McEliece cryptosystem.

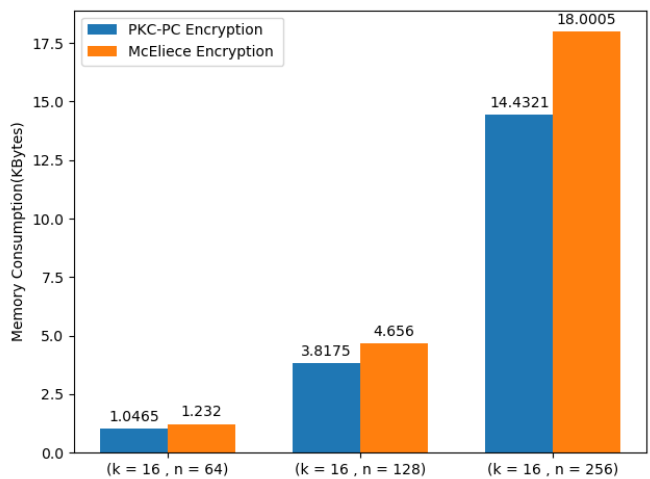

**Fig. 11:** The comparison between the encryption memory consumption (kBytes) of PKC-PC and McEliece cryptosystem implemented on Raspberry Pi3 for the same plaintext length 16bits.

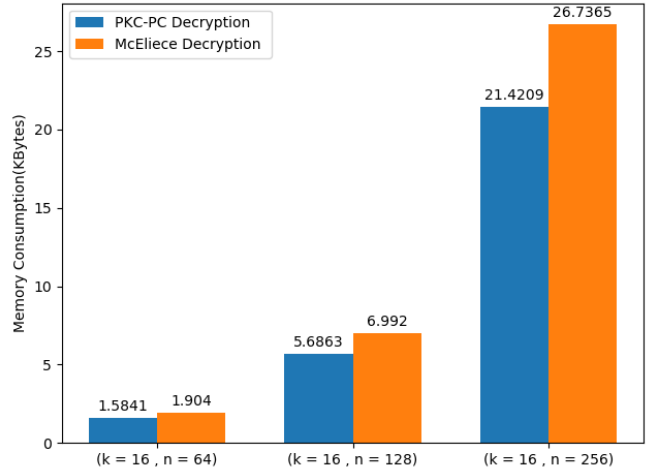

**Fig. 12:** The comparison between the decryption memory consumption (kBytes) of PKC-PC and McEliece cryptosystem implemented on Raspberry Pi3 for the same plaintext length 16bits.

### 4.3. Energy Consumption

In this research, the PKC-PC is implemented on the Raspberry Pi3 and the required current for such hardware is about 0.7A with input voltage of 1.5V. Therefore, the energy consumption in such implementation is computed as $E = I \times N \times T \times V_{cc}$, where $E$ is the energy consumption in joules, $I$ is the current consumption in amperes, $N$ is the number of clock cycles, $V_{cc}$ is the source voltage, $T$ is the clock time and $N \times T$ is the execution time. It is clear that each of the implemented cryptosystems which needs less time also consumes less energy. Figures 13 and 14 show the comparison between the energy consumption of encryption/decryption algorithms for the PKC-PC and McEliece cryptosystem implemented on the Raspberry Pi3. As shown in these figures, since the PKC-PC encryption/decryption algorithms have less execution time compared to the McEliece cryptosystem, therefore less energy consumption in the PKC-PC implementation is needed compared to the McEliece cryptosystem. However, it is clear that due to the shorter execution time of the PKC-PC encryption algorithm compared to the McEliece cryptosystem, the difference in the encryption energy consumption of such cryptosystems is greater than the difference in their decryption energy consumption. These performance parameters get decremented according to the increase in plaintext size for a constant public key, at the same time these parameters get incremented with increase in public key size.

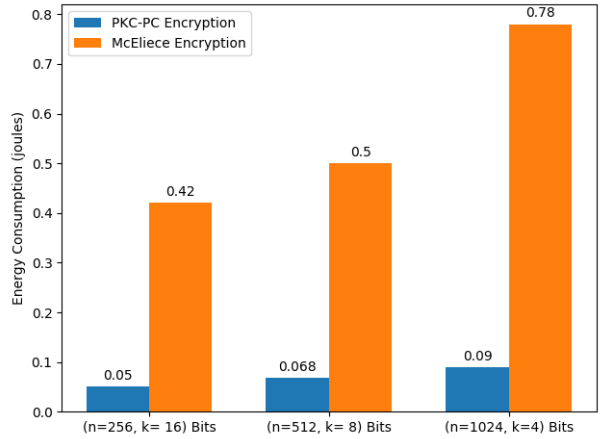

**Fig. 13:** The comparison between the encryption energy consumption (joules) of PKC-PC and McEliece cryptosystem implemented on Raspberry Pi3.

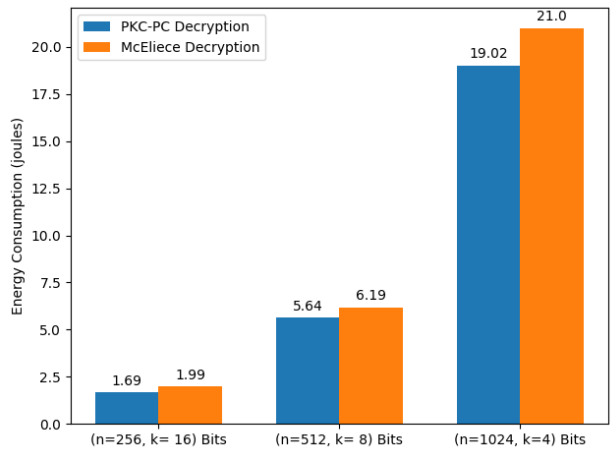

**Fig. 14:** The comparison between the decryption energy consumption (joules) of PKC-PC and McEliece cryptosystem implemented on Raspberry Pi3.

## 4.4. CPU Utilization

As shown in Table 2, the Raspberry Pi uses the ARM Cortex-A processor. According to the measurements made by the default software installed on the Raspbian operating system and assuming that $n = 256, 512, 1024$ and $k = 4, 8, 16$, the CPU utilization of implementing the encryption/decryption algorithms in the PKC-PC and McEliece cryptosystem is about 100%. Figure 15 shows the CPU utilization for PKC-PC and McEliece cryptosystem implemented on the Raspberry Pi3. As shown in this figure, implementing the PKC-PC on the Raspberry Pi3 is similar to the McEliece cryptosystem in terms of CPU utilization. In the PKC-PC implementation, during executing the key generation, encryption and decryption algorithms, the CPU temperature of Raspberry Pi3 changes between 45 and 55 degrees which can be reduced about 10 degrees by exploiting a 5-volt fan. The memory consumption is also obtained through the system's internal software which varies from 25 to 30 percent.

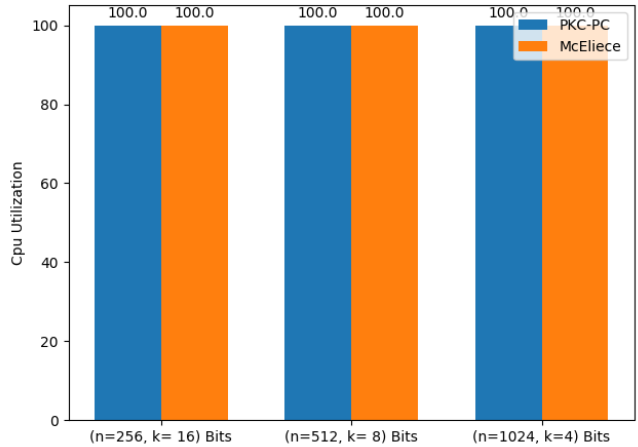

**Fig. 15:** The comparison between the CPU utilization of PKC-PC and McEliece cryptosystem implemented on Raspberry Pi3 for $n = 256, 512, 1024$ and $k = 4, 8, 16$.

## 5. Conclusion

In this paper, the hardware implementation of PKC-PC is considered. The performance parameters of such implementation, e.g. encryption/decryption execution time, energy consumption and CPU utilization have been measured and hence the implementation efficiency of PKC-PC is evaluated. Also, the measured parameters are then compared to the performance parameters of implemented McEliece public key cryptosystem. We consider two cases in such comparison, i.e. considering the same plaintext length and variable public key lengths or considering the same public key length and variable plaintext lengths. Analyses of the implementation results show that the encryption/decryption execution time, power consumption and memory consumption of PKC-PC is less than the McEliece cryptosystem for both considered cases. In the continuation of this research, it is possible to consider optimizing the successive cancellation (SC) decryption algorithm and using newer Raspberry Pi models in order to significantly reduce the encryption/decryption execution time and energy consumption of PKC-PC.